\documentstyle[12pt,epsfig]{article}

\newcommand{\be}{\begin{equation}}
\newcommand{\ee}{\end{equation}}
\def\aprle{\buildrel < \over {_{\sim}}}

\begin{document}
\topmargin 0pt
\oddsidemargin=-0.4truecm
\evensidemargin=-0.4truecm
\renewcommand{\thefootnote}{\fnsymbol{footnote}}
\newpage
\setcounter{page}{1}
\begin{titlepage}     
\vspace*{-2.0cm}
\begin{flushright}
FISIST/8-2000/CFIF \\
hep-ph/0008134
\end{flushright}
\vspace*{0.5cm}
\begin{center}
\vspace*{0.2cm}
{\Large \bf Floquet theory of neutrino oscillations in the earth}
\footnote{Contribution to the special issue of {\it Yadernaya Fizika}
dedicated to the memory of A.B. Migdal}
\\
\vspace{1.0cm}

{\large E. Kh. Akhmedov
\footnote{On leave from National Research Centre Kurchatov Institute, 
Moscow 123182, Russia. E-mail: akhmedov@cfif.ist.utl.pt}}\\
\vspace{0.05cm}
{\em Centro de F\'\i sica das Interac\c c\~oes Fundamentais (CFIF)} \\
{\em Departamento de F\'\i sica, Instituto Superior T\'ecnico}\\
{\em Av. Rovisco Pais, P-1049-001 Lisboa, Portugal}\\
\end{center}
\vglue 1.2truecm

\begin{abstract}
We review the Floquet theory of linear differential equations with 
periodic coefficients and discuss its applications to neutrino
oscillations in matter of periodically varying density. In particular, 
we consider parametric resonance in neutrino oscillations which can 
occur in such media, and discuss implications for oscillations of 
neutrinos traversing the earth and passing through the earth's core. 
\end{abstract}
\end{titlepage}
\renewcommand{\thefootnote}{\arabic{footnote}}
\setcounter{footnote}{0}
\newpage
\section{Introduction}
All oscillating systems are very much alike, and there are many
similarities between oscillating neutrinos and, e.g., pendulums or 
electromagnetic circuits. In particular, neutrino oscillations in vacuum
or in matter of constant density are analogous to oscillations of a simple 
pendulum; resonantly enhanced neutrino oscillations in a matter of 
monotonically varying density (the Mikheyev-Smirnov-Wolfenstein (MSW) effect 
\cite{MS,W}) are similar to oscillations of two weakly coupled pendulums 
of slowly changing lengths \cite{MS2,Wein}. It is therefore natural to ask 
the following question:  Are there any other resonance phenomena in mechanics 
or electromagnetism that might have analogues in neutrino physics? 

One such phenomenon is parametric resonance. The parametric resonance 
can occur in dynamical systems with time-varying parameters when 
there is a certain correlation between these variations and the values of 
the parameters themselves. Best studied is the parametric resonance in 
systems with periodically varying parameters. While the periodicity makes it
easier to satisfy the resonance conditions and also simplifies the
analysis, it is not really necessary: parametric resonance can occur even 
in stochastic systems (see, e.g., \cite{stoch}). In the present paper we 
will concentrate on the systems with periodically varying parameters. 

A textbook example of a system in which the parametric resonance can 
occur is a pendulum with vertically oscillating
point of support \cite{LL,Ar}. Under certain conditions topmost, normally
unstable, equilibrium point becomes stable. The pendulum can oscillate around 
this point in the upside-down position. Another example, familiar to
everybody, is a swing, which is just a pendulum with periodically changing 
effective length. It is the parametric resonance that makes it possible to
rock a swing.

What would be an analogue of the parametric resonance for neutrino systems? 
Since matter affects neutrino oscillations, periodically varying conditions 
can be achieved if a beam of oscillating neutrinos propagates through a 
medium with periodically modulated density. For certain relations between
the period and amplitude of density modulation and neutrino oscillation length 
and mixing angle, the parametric resonance occurs, and the oscillations can 
be strongly enhanced. The probability of neutrino transition from one flavor 
state to another may become equal to unity. This phenomenon is very different 
from the MSW effect. Indeed, at the MSW resonance the neutrino mixing in 
matter becomes maximal ($\theta_m=\pi/4$) even if the vacuum mixing angle 
$\theta_0$ is small. This leads to large-amplitude neutrino oscillations in a 
matter of constant density equal (or almost equal) to the resonance density, 
or to a strong flavor conversion in the case of matter density slowly
varying along the neutrino path and passing through the resonance value. 

The situation is quite different in the case of the parametric resonance.
The mixing angle in matter does not in general become large (there is 
no level crossing).  What happens is an amplification of the transition 
probability because of specific phase relationships. Thus, in the case of the 
parametric resonance it is the {\em phase} of oscillations (rather than their 
amplitude) that undergoes important modification. The total flavour conversion 
can take place even if the mixing angles both in vacuum {\em and} in matter 
are small.  

The possibility of the parametric resonance of neutrino oscillations   
was suggested independently in \cite{ETC} and \cite{Akh1}. In these papers
approximate solutions for sinusoidal matter density profile were found. In 
\cite{Akh1} also an exact analytic solution for the periodic step-function
(``castle wall'') density profile was obtained. Parametric effects in
neutrino oscillations were further studied in \cite{KS} where combined action 
of the parametric and MSW resonances and possible consequences for solar and 
supernova neutrinos were considered. In this paper also the stochastic 
parametric resonance in neutrino oscillations was briefly discussed. 

Although the parametric resonance in neutrino oscillations is certainly an 
interesting physical phenomenon, it requires that very special conditions 
be satisfied. Unfortunately, these conditions cannot be created in the 
laboratory because this would require either too long a baseline or neutrino 
propagation in a matter of too high a density. 
Until recently it was also unclear whether a natural object exists where these
conditions can be satisfied for any known source of neutrinos. This situation 
has changed with a very important observation by Liu and Smirnov \cite{LS} 
(see also \cite{LMS}), who have shown that the parametric resonance conditions 
can be approximately satisfied for the oscillations of atmospheric $\nu_\mu$ 
into sterile neutrinos $\nu_s$ inside the earth. 
The density profile along the trajectories of neutrinos crossing the 
earth and passing through its core (mantle-core-mantle) is to a good
approximation  a piece of the periodic step-function profile, and their
oscillations can be parametrically enhanced. 
Even though the neutrinos pass only through ``1.5 periods'' of density 
modulations (this would be exactly one period and a half if the distances 
neutrinos travel in the mantle and in the core were equal), the parametric 
effects on neutrino oscillations in the earth can be quite strong. 
Subsequently, it has been pointed out in \cite{P1} that the parametric 
resonance conditions can also be satisfied (and to even a better accuracy) 
for the $\nu_2\leftrightarrow \nu_{e}$ oscillations in the earth in the case 
of the $\nu_e$ - $\nu_{\mu(\tau)}$ mixing.  
This, in particular, may have important implications for the solar neutrino 
problem. The parametric resonance in the oscillations of solar and atmospheric 
neutrinos in the earth was further explored in a number of papers
\cite{Akh2,ADLS,CMP,Akh3,Akh4,CP,AS}. 

In the present paper we review the Floquet theory of linear differential 
equations with periodic coefficients and consider its applications to neutrino 
oscillations and, in particular, to oscillations of neutrinos inside the earth. 
The paper is organized as follows. In sec. 2 we briefly review the Floquet 
theory and its applications to the analyses of the stability of the
solutions. In sec. 3 we discuss the peculiarities of the Floquet theory 
in the case of the time dependent Schr\"odinger equations with periodic 
Hamiltonians. In sec. 4 we consider applications of the Floquet theory to
neutrino oscillations in matter of periodic step function (``castle wall'') 
density profile. In sec. 5 we review the implications of the parametric
resonance of neutrino oscillations for neutrinos traversing the earth and 
passing through its core. In the last section the results are 
discussed and the conclusions are given.

\section{Differential equations with periodic coefficients}

We shall now briefly review the Floquet theory of systems of linear
differential equations with periodic coefficients. More detailed
discussion can be found, e.g., in \cite{Ar,St}.

\subsection{Preliminaries}
Let us start with a few well known fact from the general theory of linear 
differential equations. Consider a system of $n$ homogeneous linear
differential equations 
\be
\dot{\psi} = {\cal A}(t) \psi\,,
\label{eq1}
\ee
where $\psi$ is an $n$-component column vector, $\psi=(\psi_1,...,\psi_n)^T$, 
${\cal A}(t)$ is an $n\times n$ matrix with piecewise continuous elements,
and overdot denotes differentiation with respect to $t$. 
Eq.~(\ref{eq1}) has $n$ linearly independent continuous nontrivial
solutions $\psi^{(j)}(t)$, $j=1,...,n$. From linearity of (\ref{eq1}) it 
follows that any linear combination of the solutions is also a solution. 
Any set of $n$ linearly independent solutions $\psi^{(j)}(t)$ of (\ref{eq1}) 
forms the so-called fundamental set, and a matrix whose columns are 
$\psi^{(j)}$ is called the fundamental matrix. Given an initial
condition $\psi(t_0)=\psi_0$, eq. (\ref{eq1}) has a unique solution
$\psi(t)$. Any solution of eq. (\ref{eq1}) can be represented as a linear 
combination (with constant coefficients) of the solutions forming a 
fundamental set, or equivalently as a product of a fundamental matrix 
and a constant vector. 

Let $\psi(t)$ be the solution of (\ref{eq1}) with the initial condition 
$\psi(t_0)=\psi_0$. Let us introduce the evolution matrix $U(t,t_0)$ through 
the relation 
\be
\psi(t)=U(t,t_0)\psi_0\,.
\label{U1}
\ee
{}From the definition of the evolution matrix it immediately follows that 
\be
U(t,t_0)=U(t,t_1) U(t_1,t_0)\,, \quad\quad U(t_0,t_0)=I\,,
\label{U2}
\ee
where $I$ is the $n\times n$ unit matrix. It is easy to check that the 
columns of $U(t,t_0)$ are solutions of eq. (\ref{eq1}) with the initial 
conditions $\psi^{(j)}_i(0)=\delta_{ij}$. Thus $U(t,t_0)$ is a fundamental 
matrix, and any solution of (\ref{eq1}) can be written in the form 
(\ref{U1}). The determinant of $U(t,t_0)$ is given by
\be
det[U(t,t_0)]=\exp\left\{\int_{t_0}^t tr[{\cal{A}}(t')]\,dt'\right\}\,.
\label{detU}
\ee
Eq. (\ref{detU}) follows from two facts: (1) the derivative of a determinant 
is the sum of $n$ determinants formed by replacing elements of one row of
the original matrix by their derivatives, and  (2) the columns of $U(t,t_0)$ 
are solutions of eq. (\ref{eq1}). Since the elements of ${\cal A}$
are non-singular, $det[U]$ does not vanish, i.e. the matrix $U(t,t_0)$ is 
non-singular. From (\ref{U2}) one finds 
\be
U(t_2,t_1)^{-1}=U(t_1,t_2)\,.
\label{UU3}
\ee
Without loss of generality, one can always choose the matrix ${\cal A}(t)$ 
to be traceless. Indeed, substituting 
\be
\psi(t)=e^{\alpha(t)} \psi'(t)\,, \quad \quad
\alpha(t)=\frac{1}{n}\int^t tr[{\cal{A}}(t')]\,dt'
\label{sub}
\ee
one finds that $\psi'$ satisfies 
\be
\dot{\psi}' = {\cal A}'(t) \psi'\,, \quad\quad {\cal A}'(t) = 
{\cal A}(t)-\frac{1}{n} tr[{\cal A}(t)]\,,
\label{prime}
\ee
i.e. ${\cal A}'$ is traceless. In what follows we will be always assuming 
that the transformation (\ref{sub}) has been performed, i.e. will be
considering only traceless matrices ${\cal A}$. Eq. (\ref{detU}) then
yields 
\be
det[U(t,t_0)] = 1\,.
\label{detU2} 
\ee
Notice that in general $U(t,t_0)$ is not unitary. 

\subsection{Floquet theory}

Let us now turn to the case of differential equations (\ref{eq1}) with 
periodic coefficients,
\be
{\cal A}(t+T)={\cal A}(t)\,,
\label{per1}
\ee
where $T$ is the period. Hereafter we shall be always assuming that
(\ref{per1}) is satisfied, 
without specifying this each time explicitly. From the periodicity of 
${\cal A}(t)$ it follows that if $\psi(t)$ is a solution of (\ref{eq1}), 
so is $\psi(t+T)$. Consider the solution $\psi(t)$ with the initial 
condition $\psi(s)=\psi_0$. We have $\psi(t+T)=U(t+T,s)\psi_0$. On the
other hand, since $\psi(t+T)$ is also a solution at $t$, we have 
$\psi(t+T)=U(t,t_0)\psi(t_0+T)=U(t,t_0)U(t_0+T,s)\psi_0$. Equating these 
two expressions for $\psi(t+T)$ one obtains 
\be
U(t+T,s)=U(t,t_0) U(t_0+T,s)\,.
\label{prop1} 
\ee
This is a very important property of the evolution matrix of differential 
equations (\ref{eq1}) with periodic coefficients. In particular, taking
$s=t_0+T$ and $s=t_0=0$ one obtains from (\ref{prop1}), respectively,  
\be
U(t+T,t_0+T)=U(t,t_0)\,,
\label{prop2} 
\ee
\be
~~U(t+T,0)=U(t,0)\, U(T,0)\,.
\label{prop3} 
\ee
The first of these equations means that the evolution matrix does not change
if both its arguments are shifted by the period $T$ (and, by induction, 
by any integer number $k$ of periods). From the second equality it follows, 
in particular, that the matrix of evolution over $k$ periods satisfies 
\be
U(kT,0)=U(T,0)^k\,.
\label{power}
\ee 
The matrix of evolution over one period $U(T,0)\equiv U_T$ plays a very 
important role in the theory of differential equations with periodic 
coefficients; it is called the {\em monodromy matrix}. 

As has already been pointed out, if $\psi(t)$ is a solution of (\ref{eq1}),  
so is $\psi(t+T)$. This does not in general mean that $\psi(t+T)=
\psi(t)$, i.e. the solutions of the equations with periodic coefficients
are not in general periodic. There are, however, solutions which satisfy 
\be
\psi(t+T)=\sigma \psi(t)\,, 
\label{norm1}
\ee
i.e. they are multiplied by a number when $t$ is shifted by the period. Such 
solutions are called {\em normal}\,; they play an important role in the
analysis of the stability of the solutions of eq. (\ref{eq1}).

To analyse the properties of the normal solutions, let us show that the
evolution matrix for a system of linear differential equations (\ref{eq1}) 
with periodic coefficients can be written as a product of a periodic matrix 
and an exponential matrix. Since the monodromy matrix $U_T$ is non-singular, 
it can be represented as an exponential of another matrix:
\be
U_T\equiv U(T,0)=e^{{\cal B} T}\,.
\label{exp}
\ee
Let us now show that the matrix $P(t)=U(t,0)e^{-{\cal B} t}$ is periodic. 
We have $P(t+T)=U(t+T,0) e^{-{\cal B}(t+T)}=U(t,0)U(T,0)e^{-{\cal B} T}
e^{-{\cal B} t}=U(t,0) e^{-{\cal B} t}=P(t)$, where we have used (\ref{prop3}) 
and (\ref{exp}). Thus the evolution matrix can be written as 
\be
U(t,0)=P(t)e^{{\cal B} t}\,,\quad\quad P(t+T)=P(t)\,.
\label{U4}
\ee
{}From this expression it follows, in particular, that the vector $\chi$ 
introduced through $\psi=P(t)\chi$ satisfies the differential equation 
with constant coefficients: 
\be
\dot{\chi}={\cal B}\chi\,.
\ee 

Let $\phi^{(j)}_0$ be a (constant) eigenvector of the matrix ${\cal B}$
with the eigenvalue $\alpha_jT^{-1}$. It is then also an eigenvector of
$U_T$ with the eigenvalue $\sigma_j=e^{\alpha_j}$:
\be
{\cal B}\phi^{(j)}_0=\frac{\alpha_j}{T}\phi^{(j)}_0\,; \quad \quad
U_T\phi^{(j)}_0=\sigma_j\phi^{(j)}_0=e^{\alpha_i} \phi^{(j)}_0\,.
\label{eigen}
\ee
The numbers $\sigma_j$ are called the characteristic numbers and $\alpha_j$, 
the characteristic exponents. From (\ref{prop3}), (\ref{U4}) and (\ref{eigen}) 
it follows that any normal solution $\phi^{(j)}(t)$ of eq. (\ref{eq1})
can be written as 
\be
\phi^{(j)}(t)=U(t,0)\phi^{(j)}_0=P(t) e^{\alpha_j(t/T)}\phi^{(j)}_0=
P(t) \sigma_{j}^{t/T}\phi^{(j)}_0\,.
\label{norm2}
\ee
Thus eigenvectors of $U_T$ give rise to normal solutions. The monodromy
matrix $U_T$ has $n$ eigenvalues. As follows from (\ref{detU2}), they 
satisfy
\be
\prod_{j=1}^n \sigma_j=1\,.
\label{prod}
\ee
If all the eigenvalues $\sigma_j$ are different, $U_T$ has $n$ 
linearly independent eigenvectors and therefore there are $n$ linearly 
independent normal solutions $\phi^{(j)}(t)$ of system (\ref{eq1}).
Hence, they form a fundamental set, and any solution of (\ref{eq1}) can be 
written as a linear combination of the normal solutions $\phi^{(j)}(t)$
with constant coefficients. If $U_T$ has repeated eigenvalues, the 
situation is more complicated and will be discussed below. 

It follows from eq. (\ref{norm2}) that the characteristic exponents (or 
characteristic numbers) determine the boundedness of the normal solutions,
and therefore of the general solution of eq. (\ref{eq1}). Consider, for 
example, the case $n=2$. Assume first that the characteristic exponents are 
real. The characteristic numbers $\sigma_1$ and $\sigma_2$ are then real, too. 
Eq. (\ref{prod}) gives $\sigma_1 \sigma_2=1$, and if $\sigma_1\ne \sigma_2$ 
(i.e. they are simple eigenvalues of $U_T$), the absolute value of one of
them is greater than one. It then follows from eq. (\ref{norm2}) that the 
corresponding normal solution is unbounded. If the characteristic exponents 
are purely imaginary, the characteristic numbers are of modulus one, i.e. 
$\sigma_1=\sigma_2^*$, $|\sigma_1|=1$. In this case both normal solutions 
are bounded (the matrix $P(t)$ in (\ref{norm2}), being continuous and 
periodic, is obviously bounded). If the characteristic exponents are 
complex, the normal solutions are bounded when the real parts of all the 
characteristic exponents are non-positive, while if at least one of the 
characteristic exponents has a positive real part, there are unbounded 
normal solutions. This property holds in general, i.e. for an arbitrary
$n$.   

Let us now discuss the case of repeated eigenvalues of $U_T$. Consider 
again the case $n=2$ as an example. Assume that the eigenvalues of $U_T$ 
coincide, $\sigma_1=\sigma_2\equiv \sigma$. The matrix $U_T$ in this case 
has the general form 
\footnote{Another possibility would be to have the zero element $(U_T)_{21}$ 
instead of $(U_T)_{12}$, but the corresponding matrix can be reduced to 
that in (\ref{UT1}) by a renumbering of the basis states.} 
\be
U_T=
\left(\begin{array}{cc}
\sigma & 0 \\
a & \sigma \end{array} \right)\,.
\label{UT1}
\ee
It can be readily seen that for $a=0$ the monodromy matrix $U_T$ has two 
linearly independent and orthogonal eigenvectors corresponding to the same 
eigenvlue $\sigma$; they can be taken, e.g., as $\phi^{(1)}_0=(0, 1)^T$ and 
$\phi^{(2)}_0 =(1, 0)^T$. However, for $a\ne 0$ there is only one eigenvector, 
$\phi^{(1)}_0$.  It gives rise to the normal solution $\phi^{(1)}(t)$ through 
eq. (\ref{norm2}). This normal solution can be used as one of the basis 
solutions $\psi^{(j)}(t)$ constituting a fundamental set, $\psi^{(1)}(t)= 
\phi^{(1)}(t)$. Let $\psi^{(2)}(t)$ be another solution of (\ref{eq1}), 
linearly independent from $\psi^{(1)}(t)$. Then $\psi^{(1)}(t)$ and 
$\psi^{(2)}(t)$ form a fundamental set. Since $\psi^{(j)}(t+T)$ are also 
solutions, they can be written as linear combinations of $\psi^{(1)}(t)$ 
and $\psi^{(2)}(t)$: 
\begin{eqnarray}
& & \psi^{(1)}(t+T)=\sigma \psi^{(1)}(t)\,, \nonumber \\
& & \psi^{(2)}(t+T)=a'\psi^{(1)}(t) + b'\psi^{(2)}(t)\,,
\label{rep1}
\end{eqnarray}
Using the freedom of normalization of the vectors $\phi^{(j)}_0$, one can 
always choose $a'=a$. Since $\sigma$ is a double root of the characteristic 
equation of the monodromy matrix $U_T$, $b'=\sigma$.  
\footnote{An easy way to see this is to consider eqs. (\ref{rep1}) at 
$t=0$ and use the explicit form (\ref{UT1}) of the matrix $U_T$.}
Eqs. (\ref{rep1}) can therefore be rewritten as 
\begin{eqnarray}
& & \psi^{(1)}(t+T)=\sigma \psi^{(1)}(t)\,, \label{rep2a} \\
& & \psi^{(2)}(t+T)=a \psi^{(1)}(t) + \sigma \psi^{(2)}(t)\,.
\label{rep2}
\end{eqnarray}
Let us introduce a matrix $M$ which relates $\psi^{(2)}_0$ and 
$\psi^{(1)}_0\equiv \phi^{(1)}_0$: $\psi^{(2)}_0=M \psi^{(1)}_0$ 
\footnote{The matrix $M$ is not uniquely defined, but this is unimportant 
for our purposes.}. One can now find the relation between 
$\psi^{(2)}(t)$ and $\psi^{(1)}(t)$:  
\be
\psi^{(2)}(t)=W(t)\psi^{(1)}(t)\,, \quad \quad W(t)=U(t,0) M U(t,0)^{-1}\,.
\label{Wt}
\ee
Therefore $\psi^{(2)}(t+T)=W(t+T)\psi^{(1)}(t+T)=W(t+T)\sigma\psi^{(1)}(t)$. 
On the other hand, from (\ref{rep2}) and (\ref{Wt}), $\psi^{(2)}(t+T)= 
a\psi^{(1)}(t)+\sigma \psi^{(2)}(t)=[a+\sigma W(t)]\psi^{(1)}(t)$. Equating 
these two expressions for $\psi^{(2)}(t+T)$ one finds
\be
W(t+T)=W(t)+\frac{a}{\sigma}\,.
\ee
Therefore the matrix $F(t)$ defined through
\be
F(t)=W(t)-\frac{a}{\sigma}\,\frac{t}{T}  
\label{Ft}
\ee
is periodic with the period $T$. From eqs. (\ref{Wt}) and (\ref{Ft}) we
then find 
\be
\psi^{(2)}(t)=\left[\frac{a}{\sigma}\,\frac{t}{T}+F(t)\right]
\psi^{(1)}(t)\,, \quad\quad F(t+T)=F(t)\,.
\label{rep3} 
\ee
This is the result we were looking for. Together with eq. (\ref{rep2a}) 
it states that in the case of the double roots of the characteristic 
equation of $U_T$ (i.e. in the case of coinciding characteristic exponents), 
a fundamental set can be chosen to consist of a normal solution and a
solution which is a linear combination of a linearly growing and 
periodic functions multiplied by the normal solution. Thus, if $a\ne 0$, 
there are unbounded solutions. As follows from (\ref{rep3}), if $a=0$, then   
$\psi^{(2)}(t)$, being a product of a periodic matrix and a normal 
solution, is also a normal solution. Thus in this case both solutions
forming a fundamental set can be chosen to be normal. This is 
in accord with the fact that for $a=0$ the matrix $U_T$ in eq.~(\ref{UT1}) 
has two linearly independent eigenvectors. 

It is instructive to see how the linear growth of $\psi^{(2)}(t)$ arises 
from the general expression $\psi^{(2)}(t)=P(t)e^{{\cal B} t} \psi^{(2)}_0$.
{}From (\ref{exp}) and (\ref{UT1}) one finds
\be
{\cal B } T =
\ln U_T=
\ln[\sigma (I+\Delta)]
= \ln\sigma +\ln(I+\Delta)=
\ln\sigma +\Delta-\frac{\Delta^2}{2}+
\frac{\Delta^3}{3}-...
\,,
\label{B1}
\ee
where 
$$\Delta=\left(\begin{array}{cc}
0 & 0 \\
a/\sigma & 0 \end{array} \right)\,.$$
Since $\Delta^2=0$, eq. (\ref{B1}) gives ${\cal B}T= \ln\sigma+\Delta$, 
hence 
\be
e^{{\cal B}t}=\sigma^{t/T}e^{\Delta (t/T)}=\sigma^{t/T}(I+\frac{t}{T}
\Delta)\,. 
\label{exp2}
\ee
Notice that the matrix $\Delta$ annihilates $ \phi^{(1)}_0$, and therefore
(\ref{exp2}) does not contradict $\psi^{(1)}(t)$ 
being a normal solution.  Using eq. (\ref{exp2}) one can find a simple 
representation of the matrix $F(t)$ entering into eq. (\ref{rep3}), 
\be
F(t)=P(t) M P(t)^{-1}\,,
\ee
from which the periodicity of $F(t)$ is obvious. 

The above result can be generalized to the case $n>2$. If the characteristic 
equation of the monodromy matrix $U_T$ has repeated roots (i.e. some of the 
characteristic numbers coincide), a fundamental set can be chosen to 
consist of normal solutions and solutions which are linear combinations of 
polynomials in $t$ and periodic matrices multiplied by normal solutions. 

We have seen that in the case when some of the characteristic exponents 
have positive real parts, there are exponentially growing solutions, while 
if some of the characteristic exponents coincide, there are in general 
polynomially growing solutions. The existence of such unbounded solutions 
signifies instabilities due to the parametric resonance. 

It should be noticed that instead of considering a system of $n$ first 
order linear equations with periodic coefficients (\ref{eq1}) one could 
equivalently consider one equation of the $n$th order. In particular, in
the case $n=2$, a general second order equation with periodic coefficients
is obtained, which is called the Hill equation. A pendulum with vertically 
oscillating point of support mentioned in the Introduction is described by 
this equation. If the oscillations of the point of support are harmonic, the 
pendulum is described by the well known Mathieu equation. In the limit of 
small-amplitude oscillations of the point of support, the instability 
condition (the condition of exponential growth of the deviation from the 
equilibrium) is \cite{LL,Ar}
\be
\Omega \equiv \frac{2\pi}{T}=\frac{2\omega}{k}\,,
\label{res1}
\ee
where $\omega$ is the frequency of the oscillations of the pendulum in the
absence of the motion of its point of support, and $k$ is an integer.
Eq. (\ref{res1}) relates the frequency of the oscillations of the point of 
support $\Omega$ at which the parametric resonance occurs 
to the oscillator eigenfrequency. In general, when the amplitude of the 
oscillations of the point of support is not small, the parametric
resonance condition depends not only on the frequency of these oscillations 
but also on their amplitude. 
In this case there are resonance regions of parameters rather than 
resonance values \cite{Ar,St}.  

In real physical systems all parameters are, of course, finite; unboundedness 
of certain solutions of eq. (\ref{eq1}) with periodic coefficients just 
reflects the fact that in general the dynamics of real systems is only 
approximately described by linear equations. For large deviations from 
equilibrium, nonlinear effects become important and eqs. (\ref{eq1}) have to 
be modified. There are, however, cases when the solutions are always bounded 
even in the linear regime, and the description by linear equations can be 
exact. Nevertheless, parametric resonance is possible in such systems as well. 
One example of such a situation is given by Schr\"odinger equations with 
periodic Hamiltonians,  which we discuss next. 

\section{Schr\"odinger equations with periodic Hamiltonians}

If the matrix ${\cal A}(t)$ in eq. (\ref{eq1}) is anti-Hermitian, the system 
of equations (\ref{eq1}), (\ref{per1}) can be written as a Schr\"odinger 
equation with a periodic Hermitian Hamiltonian:
\be
i\dot{\psi}={\cal H}(t)\psi\,, \quad\quad {\cal H}(t)^\dagger={\cal H}(t)\,, 
\quad\quad {\cal H}(t+T)={\cal H}(t)\,,
\label{Sch}
\ee
where ${\cal H}(t)=i{\cal A}(t)$. In the case of constant ${\cal H}$, 
eq. (\ref{Sch}) describes oscillations between the components of
$\psi$ characterized by $n-1$, in general different, frequencies (out of
$n$ eigenvalues of ${\cal H}$ only $n-1$ are independent since $tr{\cal
H}=0$). In particular, spin precession in a constant magnetic field or  
neutrino oscillations in vacuum or in matter of constant density are
described by such an equation. Eq. (\ref{Sch}) with time-dependent periodic 
Hamiltonians describes many physical systems, e.g., atoms in a laser field or 
electron paramagnetic resonance. 
It also describes neutrino oscillations in a medium of periodically
modulated density. 
 
Because of the Hermiticity of ${\cal H}(t)$, the evolution matrix $U(t,t_0)$ 
of eq. (\ref{Sch}) is unitary, i.e. the norm of the vector $\psi$ is 
conserved. Therefore all the solutions of eq. (\ref{Sch}) are bounded. 
The parametric resonance in the systems described by eq.~(\ref{Sch}) has 
therefore some peculiarities, which we shall discuss below.

One consequence of the unitarity of the evolution matrix $U(t,t_0)$ is 
that all the characteristic exponents are purely imaginary. In addition, 
since the polynomial growth of the solutions is not allowed in this case, 
even in the case of repeated roots of the characteristic equations of the
monodromy matrix $U_T$ there are $n$ linearly independent normal solutions 
which form a fundamental set. This actually directly follows from the 
fact that a unitary $n\times n$ matrix has exactly $n$ linearly independent 
eigenvectors, irrespective of whether or not all the roots of its 
characteristic equations are simple.  

We shall now discuss general properties of the solutions of eq. (\ref{Sch}),
using again the case $n=2$ as an example. The results, in particular, will 
apply to the problem of two-flavour neutrino oscillations in a matter of
periodically modulated density. 

Let us start with a few general remarks about the solutions of the 
Shr\"odinger equation with time dependent (but not necessarily periodic) 
Hamiltonian in the case $n=2$. First, we notice that without loss of 
generality the Hamiltonian ${\cal H}(t)$ can be considered to be real. 
Indeed, in the case of a complex Hamiltonian, a rephasing of the components 
$\psi_{1,2}$ of $\psi$ by the factors $\exp(\pm i\beta(t)/2)$ where 
$\beta(t)=arg[{\cal H}_{12}(t)]$ transforms the Hamiltonian to the form 
\be
{\cal H}(t)=
\left(\begin{array}{cc}
-A(t) & B(t) \\
B(t) & 
A(t)  
\end{array} \right)
\label{H1}
\ee
with real $A(t)$ and $B(t)$. Notice that this rephasing transformation 
preserves the trace of the Hamiltonian. 

Next, we notice that the Hamiltonian ${\cal H}(t)$ in (\ref{H1}) can be 
written as 
\be
{\cal H}(t)=B(t)\sigma_1-A(t)\sigma_3\,,
\label{H2}
\ee
where $\sigma_i$ are the Pauli matrices. Thus ${\cal H}(t)$ anticommutes 
with $\sigma_2$. From this fact it immediately follows that if $\psi(t)=
(\psi_1(t), \psi_2(t))^T$ is a solution of the Schr\"odinger equation, so 
is $\tilde{\psi}(t)=-i\sigma_2 \psi(t)^*=(-\psi_2(t)^*, \psi_1(t)^*)^T$. It 
is easy to see that $\psi(t)$ and $\tilde{\psi}(t)$ are orthogonal and 
therefore linearly independent. Thus, if one nontrivial solution of the 
Schr\"odinger equation in the case $n=2$ is known, it automatically gives 
another nontrivial solution, linearly independent from the original one. The
solutions $\psi$ and $\tilde{\psi}$ form a fundamental set and therefore 
knowledge of a single nontrivial solution of the Schr\"odinger equation
allows one to obtain the general solution. 

We now turn to the study of the Schr\"odinger equation with periodic
coefficients in the case $n=2$. Since the monodromy matrix 
$U_T=\exp({\cal B}T)$ is a unitary $2\times 2$ matrix, it can be written as 
\be
U_T=Y - i\mbox{\boldmath $\sigma$} {\bf X}=
\exp[-i(\mbox{\boldmath $\sigma$}{\bf \hat{X}}) \Phi]\, 
\label{UT2}
\ee 
with real parameters ${\bf X}$ and $Y$ (or ${\bf\hat{X}}$ and $\Phi$) 
\footnote{The reality of these parameters is a consequence of the relation
$\sigma_2 U(t,t_0)^* \sigma_2=U(t,t_0)$ which, in turn, follows from the 
properties of the Hamiltonian ${\cal H}(t)^*={\cal H}(t)$ and 
$\{{\cal H}(t), \sigma_2\}=0$.}  
which satisfy 
\be
Y^2+{\bf X}^2=1\,,\quad\quad \cos\Phi=Y\,,\quad\quad \sin\Phi=|{\bf X}|\,;
\quad\quad 
{\bf \hat{X}}\equiv \frac{{\bf X}}{|{\bf X}|}\,.
\label{param1}
\ee 
It should be noted that the form of the monodromy matrix in eq. (\ref{UT2}) 
is quite general, i.e. it does not depend on the particular form of the
functional dependence of $A(t)$ and $B(t)$, whereas the values of the 
parameters $Y$ and ${\bf X}=\{X_1, X_2, X_3\}$ (or $\Phi$ and 
${\bf \hat{X}}=\{\hat{X}_1, \hat{X}_2, \hat{X}_3\}$) are, of course,
determined by this functional dependence. 
Using (\ref{U4}) one can write the evolution matrix as
\be
U(t,0)=P(t) e^{-i(\mbox{\boldmath $\sigma$}{\bf \hat{X}}) \Phi \,(t/T)}\,,
\quad\quad\quad P(t+T)=P(t)\,. 
\label{U5}
\ee
The matrix of evolution over an integer number of periods is 
\be
U(kT,0)=(U_T)^k=
\exp[-i(\mbox{\boldmath $\sigma$}{\bf \hat{X}}) k\Phi]\,. 
\label{UkT}
\ee

{}From eqs. (\ref{exp}) and (\ref{UT2}) one finds  
\be
{\cal B}T=-i(\mbox{\boldmath $\sigma$}{\bf \hat{X}}) \Phi\,.
\label{BT}
\ee
Since ${\bf \hat{X}}$ is a unit vector, the eigenvalues of the matrix
$\mbox{\boldmath $\sigma$}
{\bf \hat{X}}$ are $\pm 1$, hence the characteristic exponents $\alpha_{1,2}$ are purely
imaginary, and the characteristic numbers $\sigma_{1,2}$ are of modulus
one:
\be
\alpha_{1,2}=\pm i\Phi\,, \quad\quad \sigma_{1,2}=e^{\pm i\Phi}\,.
\label{char}
\ee

Let us find the normal solutions. The eigenvectors $\phi^{(1,2)}_0$ of the 
monodromy matrix coincide with the eigenvectors of the matrix 
$\mbox{\boldmath $\sigma$}{\bf \hat{X}}$; they can be written as 
\be
\phi^{(1)}_0=\frac{1}{\sqrt{2}}
\left(\begin{array}{l}
\vspace*{0.15cm}
\sqrt{1+\hat{X}_3} \\
\sqrt{1-\hat{X}_3} \,e^{i\delta}
\end{array}\right)\,, \quad\quad\quad
\phi^{(2)}_0=\frac{1}{\sqrt{2}}
\left(\begin{array}{l}
\vspace*{0.15cm}
-\sqrt{1-\hat{X}_3} \,e^{-i\delta}\\
~~\sqrt{1+\hat{X}_3} 
\end{array}\right)\,, 
\label{eigen2}
\ee
where
\be
\delta=arg(\hat{X}_1+i\hat{X}_2)\,.
\label{delta}
\ee
Notice that $\phi^{(2)}_0=\tilde{\phi}^{(1)}_0\equiv -i\sigma_2
(\phi^{(1)}_0)^*$. 
The normal solutions are now found as $\phi^{(1,2)}(t)=
U(t,0)\phi^{(1,2)}_0$, which gives 
\be
\phi^{(1)}(t)=P(t)e^{-i\Phi (t/T)} \phi^{(1)}_0\,, 
\quad\quad\phi^{(2)}(t)=P(t)e^{i\Phi (t/T)} \phi^{(2)}_0\,. 
\label{norm3}
\ee
They form a fundamental set, which means that an arbitrary solution $\psi(t)$ 
can be represented as a linear combination of $\phi^{(1}(t)$ and 
$\phi^{(1}(t)$ with constant coefficients:
\be
\psi(t)=P(t)[C_1\, e^{-i\Phi (t/T)} \phi^{(1)}_0 + 
C_2 \,e^{i\Phi (t/T)} \phi^{(2)}_0]\,.
\label{decomp} 
\ee
The normal solutions (\ref{norm3}) are the products of the periodic functions 
with periods $T$ and $\tau=(2\pi/\Phi)T$. 
Thus, the general solution 
(\ref{decomp}) describes modulated oscillations between the components of 
$\psi(t)$ -- the parametric oscillations \cite{AS}. 

Let us now discuss the parametric resonance in the system under consideration. 
In the general case of eq. (\ref{eq1}), the parametric resonance typically 
corresponds to the situations when a solution becomes unbounded, i.e.
there are some values of $t$ for which the modulus of a component of $\psi$ 
can exceed any pre-assigned number, however large. A characteristic feature of 
the parametric resonance is that this can happen even for arbitrarily small 
amplitude of variations of the coefficients of eq. (\ref{eq1}). 

In the case of the systems described by eq. (\ref{Sch}), the parametric 
resonance corresponds to a situation when there exist values of $t$ for 
which the modulus of a component of a solution can reach maximal allowed by 
unitarity value, unattainable in the case of the corresponding equation with 
constant coefficients. This can happen even for arbitrarily small amplitude 
of the variations of the coefficients in eq. (\ref{Sch}). Notice that the 
parametric resonance is undefined in the cases when eq. (\ref{Sch}) with 
constant coefficients itself leads to maximal amplitude oscillations.

To be more specific, consider 
the case $n=2$. If the Hamiltonian of the system was constant, 
the Schr\"odinger equation (\ref{Sch}) would describe the oscillations 
between the components 
of $\psi$ with the frequency $\omega_0=\sqrt{A^2+B^2}$ and amplitude 
$\sin^2 2\theta_0=B^2/(A^2+B^2)$. If $A \ne 0$, this amplitude is always 
less than unity. We shall now consider the case of time dependent $A$
and $B$, but will be assuming that $A(t)$ never vanishes during the period
of evolution of interest 
\footnote{In the case of neutrino oscillations in vacuum, the condition 
$A=0$ corresponds to maximum mixing, while for neutrino oscillation in 
a matter of varying density $A(t)=0$ corresponds to the MSW resonance.}. 
The transition probability in the ``quasi time-independent'' (adiabatic) 
regime then never exceeds $max\{B(t)^2/(A(t)^2+B(t)^2)\}<1$. The parametric  
resonance occurs when there are values of $t$ for which the modulus of a 
component of $\psi$, which was initially equal to zero, can reach
the maximal allowed by unitarity value, corresponding to the transition 
probability equal to one. 
This can happen even if $max\{B(t)^2/(A(t)^2+B(t)^2)\}\ll1$ and for 
arbitrarily small amplitudes of time variations of $A(t)$ and $B(t)$. 

Let us find the parametric resonance condition. 
Assume that the initial state at $t=0$ is $\psi_0=(1,0)^T$. Then at $t=kT$ 
we find
\be
\psi(kT)=U(kT,0)\psi_0=
\left(\begin{array}{c}
\vspace*{0.15cm}
\cos k\Phi-i\hat{X}_3\sin k\Phi \\
-i e^{-i\delta} \sqrt{1-\hat{X}_3^2} \,\sin k\Phi
\end{array}\right)\,.
\label{kT}
\ee
We shall now show that the parametric resonance condition is \cite{Akh2} 
\be
\hat{X}_3=0\,.
\label{paramres}
\ee
Indeed, the transition probability reaches its maximum possible value, equal 
to one, when the survival probability $|\psi_1(t)|^2$ vanishes. From 
(\ref{kT}), in the case $\hat{X}_3=0$, one has $\psi_1(t=kT)=\cos k\Phi$.
The component $\psi_1(t)$ at $t=(k+1)T$ is $\cos (k+1)\Phi$. It is easy to 
see that for an arbitrary nonzero value of $\Phi$ there is a value of $k$
for which $\cos k\Phi\le 0$, $\cos (k+1)\Phi>0$ or vice versa. Since all the 
solutions of eq. (\ref{Sch}) with regular coefficients are continuous, this 
means that there is a value $t_1$, $kT\le t_1 <(k+1)T$, for which 
$\psi_1(t_1)=0$ and the survival probability vanishes, i.e. the 
component $\psi_2$ saturates the unitarity limit. 
Thus (\ref{paramres}) is the parametric resonance condition. 

\section{Neutrino oscillations in matter with ``castle wall'' density profile} 

We shall now consider applications of the Floquet theory reviewed in the 
preceding sections to neutrino oscillations in a matter of periodically 
modulated density. In particular, we shall be interested in the
parametric resonance in neutrino oscillations. 

Consider oscillations in a two-flavour neutrino system. The evolution of the 
system in the flavour eigenstate basis is described by the Schr\"odinger
equation with the Hamiltonian (\ref{H1}) in which the parameters $A$ and $B$ 
are given by 
\be
A(t)=\frac{\Delta m^2}{4E}\cos 2\theta_0-\frac{G_F}{\sqrt{2}}\,
N(t)\,,
\quad\quad
B=\frac{\Delta m^2}{4E}\sin 2\theta_0\,,
\label{AB}
\ee
Here $G_F$ is the Fermi constant, $E$ is neutrino energy, $\Delta
m^2=m_2^2-m_1^2$, where $m_{1,2}$ are the neutrino mass eigenvalues, and
$\theta_0$ is the mixing angle in vacuum. The effective density $N(t)$
depends on the type of the neutrinos taking part in the oscillations:
\be
N=\left\{\begin{array}{lll}
N_e & {\mbox {\rm for}} &
\nu_e\leftrightarrow \nu_{\mu,\tau} \\
0 & {\mbox {\rm for}} &
\nu_\mu\leftrightarrow \nu_{\tau} \\
N_e-N_n/2 & {\mbox {\rm for}} &
\nu_e\leftrightarrow \nu_{s} \\
-N_n/2 & {\mbox {\rm for}} &
\nu_{\mu,\tau}\leftrightarrow \nu_{s}\,.
\end{array} \right.
\label{N}
\ee
Here $N_e$ and $N_n$ are the electron and neutron number densities,
respectively.
For transitions between antineutrinos one should
substitute $-N$ for $N$ in eq. (\ref{AB}). If overall matter density
and/or chemical composition varies along the neutrino path, the effective
density $N$ depends on the neutrino coordinate $t$. The instantaneous
oscillation length $l_m(t)$ and mixing angle $\theta_m(t)$ in matter are
given by
\be
l_m(t)=\pi/\omega(t)\,,\quad\quad \sin 2\theta_m(t)=B/\omega(t)\,,
\quad\quad \omega(t)\equiv\sqrt{B^2+A(t)^2}\,.
\label{matter}
\ee
The MSW resonance corresponds to $A(t_{res})=0$, $\sin
2\theta_m(t_{res})=1$.

For the parametric resonance to occur, the exact shape of the matter density 
profile is not very important; what is important is that the change in the 
density be synchronized in a certain way with the change of the oscillation 
phase. In particular, in \cite{ETC,Akh1} the case of the sinusoidal density 
profile was considered in which the neutrino evolution equation reduces to a 
modified Mathieu equation. In \cite{Akh1} the parametric resonance was
also considered for neutrino oscillations in a matter with a periodic step 
function density profile, which allows a very simple exact analytic solution. 
This solution was studied in detail in \cite{Akh2,AS}. Here we shall
review this solution and its main features. 

Consider the case when the effective density $N(t)$ (and
therefore $A(t)$) is a periodic step function:
\[
N(t)=\left\{\begin{array}{ll}N_1\,, & 0 \le t <T_1 \\
N_2\,, & T_1 \le t <T_1+T_2
\end{array} \right.
\]
\be
N(t+T)=N(t)\,,\quad T=T_1+T_2\,.
\label{A}
\ee
Here $N_1$ and $N_2$ are constants. We shall call this the ``castle wall''
density profile (see fig. 2). The function $A(t)$ is expressed by
 similar formula with constants $A_1$ and $A_2$. Thus, the Hamiltonian
${\cal H}(t)$ is also a periodic function of time with the period $T$. 
Let us denote
\be
\delta=\frac{\Delta m ^2}{4E}\,,\quad\quad V_i=\frac{G_F}{\sqrt{2}}\,
N_i\quad\quad (i=1,~2)\,.
\label{not}
\ee
In this notation
\be
A_i=\cos 2\theta_0\,\delta-V_i\,,\quad\quad
B=\sin 2\theta_0\,\delta\,,\quad\quad
\omega_i=\sqrt{(\cos 2\theta_0\,\delta-V_i)^2+
(\sin 2\theta_0\,\delta)^2}\,.
\label{AB1}
\ee

Any instant of time in the evolution of the neutrino system belongs to
one of the two kinds of the time intervals:
\be
\begin{array}{ll}
(1): & 0+nT \le t <T_1+nT \\
(2): & T_1+nT \le t <T_1+T_2+nT\,, \quad n=0,~1,~2,...
\end{array}
\label{int}
\ee
In either of these time intervals the Hamiltonian ${\cal H}$ is a constant
matrix which we denote ${\cal H}_1$ and ${\cal H}_2$, respectively.
Let us define the evolution matrices for the intervals of time $(0, T_1)$
and $(T_1, T_1+T_2)$:
\be
U_1=\exp(-i{\cal H}_1 T_1)\,, \quad\quad U_2=\exp(-i{\cal H}_2 T_2)
\label{U1U2}
\ee
The monodromy matrix is then 
\be
U_T=U_2 U_1\,.
\label{UTT1}
\ee
Let us introduce the unit vectors
\begin{eqnarray}
{\bf n}_1=\frac{1}{\omega_1}(B,~0,~-A_1)
=(\sin 2\theta_1,~0,~-\cos 2\theta_1)\,, \nonumber \\
{\bf n}_2=\frac{1}{\omega_2}(B,~0,~-A_2)
=(\sin 2\theta_2,~0,~-\cos 2\theta_2)\,,
\label{n1n2}
\end{eqnarray}
where
$\theta_{1,2}$ are the mixing angles in matter at densities $N_1$ and
$N_2$: $\theta_1=\theta_m(N_1)$, $\theta_2=\theta_m(N_2)$.
Then one can write
\be
{\cal H}_i=\omega_i (\mbox{\boldmath $\sigma$} {\bf n}_i)\,.
\label{Hi}
\ee
Using eqs. (\ref{U1U2}) - (\ref{Hi}) one can obtain the monodromy  matrix
in the form (\ref{UT2}) with the parameters $Y$ and ${\bf X}$ given by 
\be
Y=c_1 c_2-({\bf n_1 n_2}) s_1 s_2\,,   
\label{Y}
\ee
\be
{\bf X}=s_1 c_2\,{\bf n}_1+s_2 c_1\,{\bf n}_2-s_1 s_2\,
({\bf n}_1\times {\bf n}_2)\,,
\label{X}  
\ee
where we have used the notation
\be
s_i=\sin \phi_i\,,\quad c_i=\cos \phi_i\,,\quad \phi_i=\omega_i T_i\,
\quad (i=1,~2)\,.
\label{sici}
\ee
Notice that the difference of the neutrino eigenenergies in a matter of 
density $N_i$ is $2\omega_i$, so that 
$2\phi_1$ and $2\phi_2$ are the oscillations phases acquired over the 
intervals $T_1$ and $T_2$. The evolution matrix for $k$ periods 
($k$=0, 1, 2,...) is given by eq. (\ref{UkT}). 

The vector ${\bf X}$ can be written in components as
\be
{\bf X}=\left\{(s_1 c_2 \sin 2\theta_1+s_2 c_1 \sin 2\theta_2)\,,~~
-s_1 s_2 \sin(2\theta_1-2\theta_2)\,,~~
-(s_1 c_2 \cos 2\theta_1+s_2 c_1 \cos 2\theta_2) \right\}\,.
\label{comp}
\ee

Eqs. (\ref{UT2}) - (\ref{UkT}) and (\ref{Y}) - (\ref{comp}) give the exact
solution of the evolution equation for any instant of time that is  
an integer multiple of the period $T$. In order to obtain the solution for
$kT<t<(k+1)T$ one has to evolve the solution at $t=kT$ by applying the 
evolution matrix
\be
U_1(t,~kT)=\exp[-i{\cal H}_1\cdot(t-kT)]
\label{U1t}
\ee
for $kT<t<kT+T_1$ or
\be
U_2(t,~kT+T_1) U_1=\exp[-i{\cal H}_2\cdot(t-kT-T_1)] \exp[-i{\cal H}_1 T_1]
\label{U2t}
\ee
for $kT+T_1 \le t < (k+1)T$, with ${\cal H}_{1,2}$ given by eq. (\ref{Hi}).

\subsection{Parametric resonance}
Assume that the initial neutrino state at $t=0$ is a flavor eigenstate
$\nu_a$. The probability of finding another flavor eigenstate $\nu_b$ at 
a time $t>0$ (transition probability) is then $P(\nu_a\to \nu_b,~t)=
|U_{21}(t,0)|^2$. As pointed out in sec. 3, the evolution of neutrino  
system in a matter of periodically varying density has a character of 
parametric oscillations -- modulated oscillations characterized by two  
periods, $T$ and $\tau=(2\pi/\Phi)T$. 
As can bee seen from eq.~(\ref{kT}), 
the transition probability after 
passing $k$ periods of density modulation is \cite{Akh2}
\be
P(\nu_a\to \nu_b,~t=kT)=(1-\hat{X}_3^2)\sin^2\Phi_p\,, 
\quad\quad\quad \Phi_p=k\Phi\,,
\label{prob1}
\ee
where $\Phi$ was defined in (\ref{param1}). 
Notice that this expression is valid for any periodic matter density
profile, irrespective of its shape. The values of $\hat{X}_3$ and $\Phi$, of
course, depend on this shape. For neutrino oscillations in matter with the 
``castle wall'' density profile, eq.~(\ref{prob1}) corresponds to the 
evolution over an even number of layers of constant density.  
The transition probability after passing an odd number of alternating
layers, which can be considered as $k$ periods plus one additional layer
of density $N_1$ (the corresponding evolution time $t=kT+T_1$), is also
given by eq. (\ref{prob1}), the only difference being that the phase is 
now \cite{AS} 
\be
\Phi_p=k\Phi+\varphi\,
\label{Phip}
\ee
with 
\begin{eqnarray}
& & \sin\varphi = s_1 \sin 2\theta_1/\sqrt{1-\hat{X}_3^2}\,,
\nonumber \\
& & \cos\varphi = (s_1 \sin 2\theta_1 Y+s_2 \sin 2\theta_2)/
\sqrt{{\bf X}^2-X_3^2}
\label{varphi}
\end{eqnarray}
Eqs. (\ref{prob1}) - (\ref{varphi}) give the transition
probability at the borders of the layers.

\begin{samepage}
\begin{figure}
\setlength{\unitlength}{1cm}
\begin{center}
\vspace{-1.2cm}
\epsfig{file=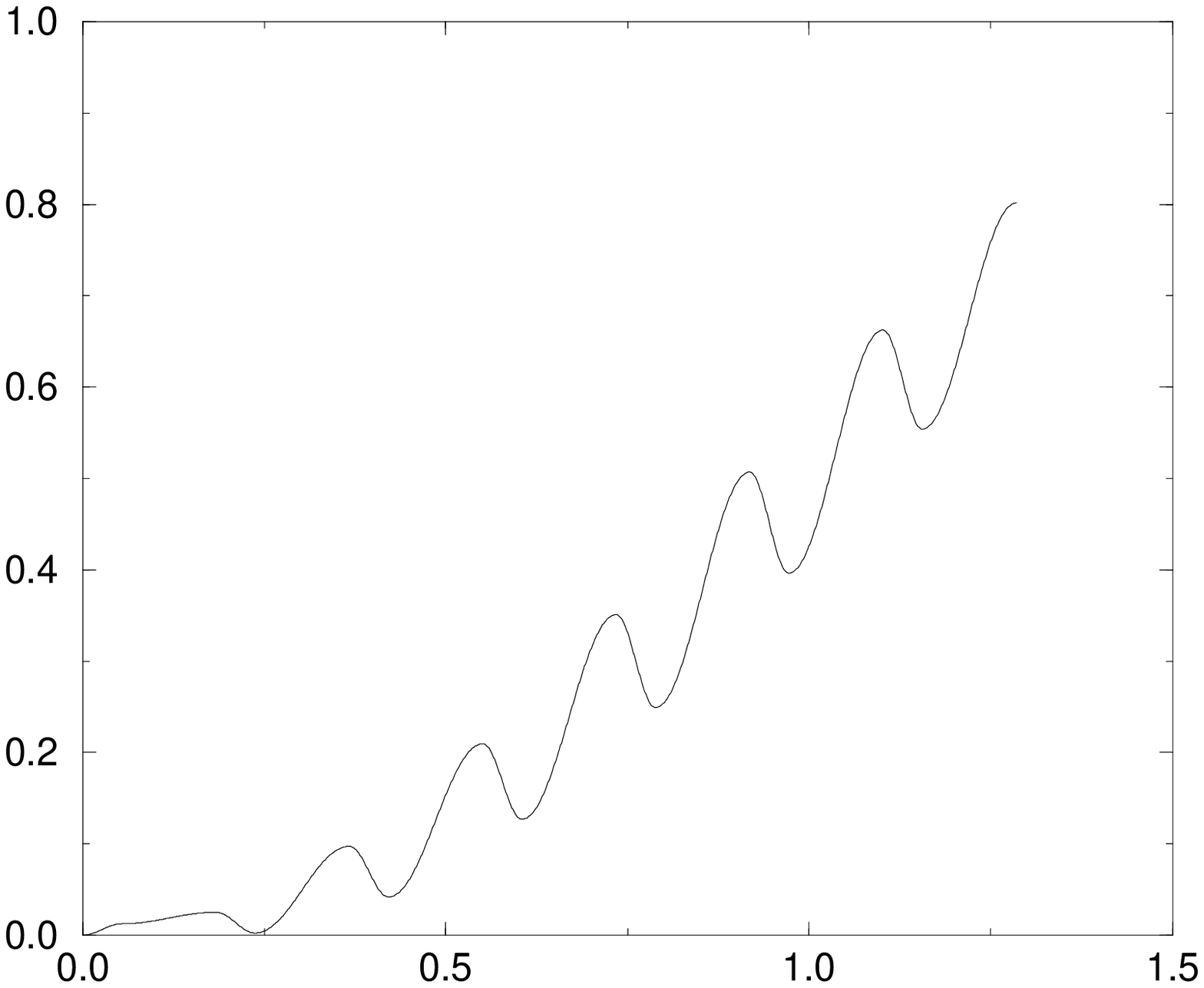,width=12cm}
\end{center}
\vspace{-1.4cm}   
\caption{\label{f1}
\small 
Coordinate dependence of the neutrino flavor transition probability 
$P$ in a matter with the ``castle wall'' density profile. 
$\sin^2 2\theta_{0}=0.01$, 
$\delta= 10^{-12}$ eV, 
$V_1=10^{-13}$ eV, $V_2=6.33\times 10^{-13}$ eV, 
$T_1=5.4\times 10^{-2}$, $T_2=0.1296$, 
all distances are in units of $R=3.23 \times 10^{13}$ eV$^{-1}$. }
\begin{center}
\epsfig{file=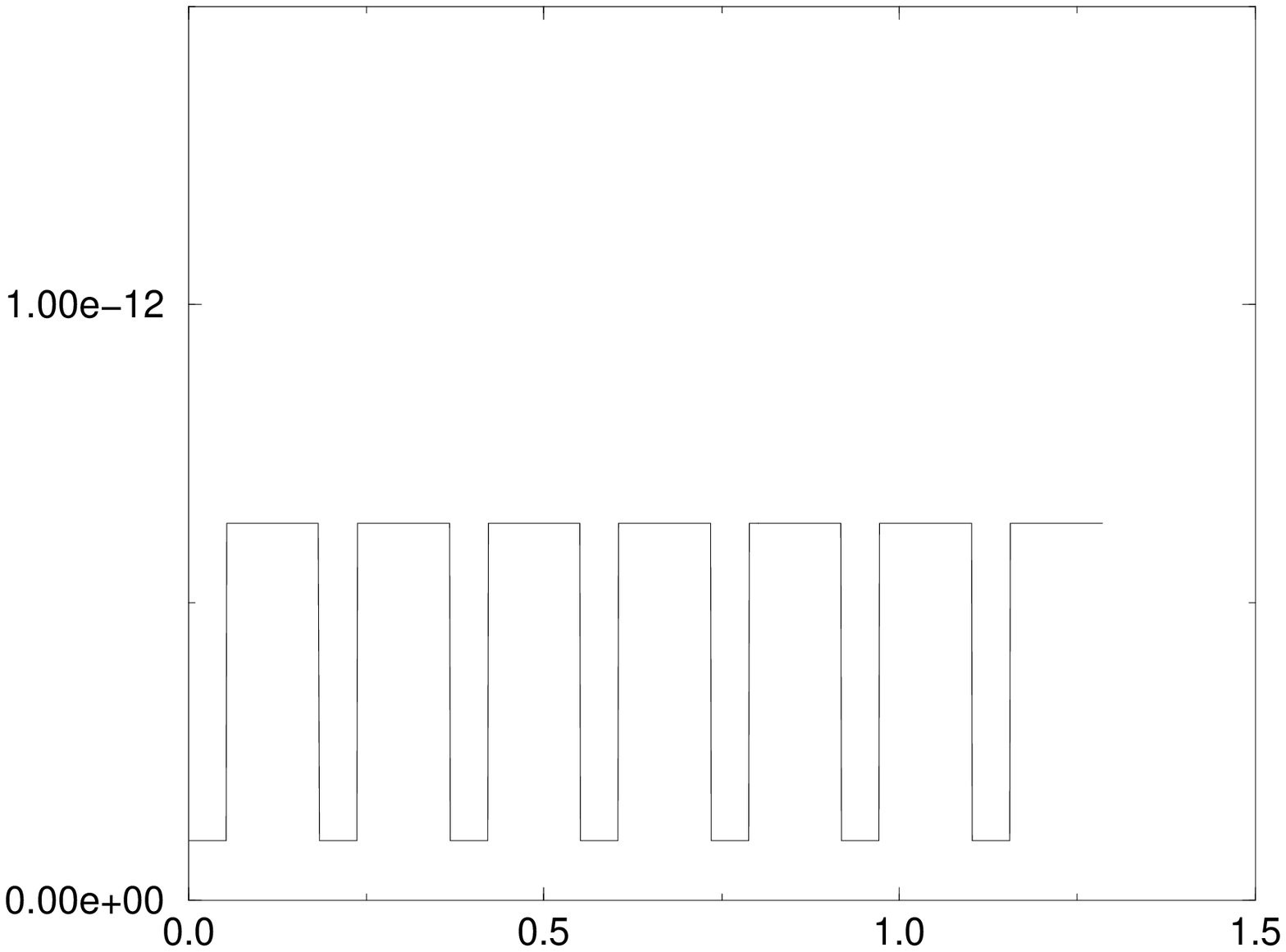,width=12cm}
\end{center}
\vspace{-1.4cm}   
\caption{\label{f2}
\small Coordinate dependence of the matter-induced neutrino potential 
[$G_F/\sqrt{2}\times$ (density profile)] for the case shown in fig. 1.}
\end{figure}
\end{samepage}  

The parametric resonance occurs when the depth of the parametric oscillations 
described by (\ref{prob1}) becomes equal to unity, i.e.  when $\hat{X}_3=0$. 
This coincides with the condition (\ref{paramres}) found in sec. 3. 
For the case of the ``castle wall'' density profile under consideration it 
reads \cite{Akh2} (see eq. (\ref{comp})):  
\be
X_3 \equiv -(s_1 c_2 \cos 2\theta_1 +s_2 c_1 \cos 2\theta_2 )=0\,.
\label{res2}
\ee
As follows from (\ref{prob1}), the maximum transition probability $P=1$
can be achieved at the borders of the layers provided that
\be
\Phi_p=\frac{\pi}{2}+n\pi\,, \quad\quad n=0, 1, 2, ...
\label{phase}
\ee

The parametric resonance condition (\ref{res2}) can be realized in two 
different ways. One possibility is that both terms on the right hand side of 
eq. (\ref{res2}) vanish. This requires $c_1 = c_2  = 0$ 
[8 -- 14], or 
\footnote{In refs. \cite{ETC,Akh1,KS} these conditions were derived for 
the particular case $k'=k''$, which includes the most important principal
resonance with $k'=k''=0$.}
\be
\phi_1=\frac{\pi}{2}+k'\pi\,, \quad\quad \phi_2=\frac{\pi}{2}+k''\pi\,,
\quad\quad k', k''=0,1,2,...
\label{rescond}
\ee
Notice that another option, $s_1=s_2=0$, leads to a trivial 
case ${\bf X}=0$, $Y=\pm 1$, in which the monodromy matrix coincides (up
to the sign) with the unit matrix, and the transition probability on the 
borders of the layers vanishes 
\footnote{We do not consider the trivial cases of the MSW resonance for
which $X_3 = 0$ because $\cos 2\theta_{i} = 0$ and $s_{i} = \pm 1$, $i = 1$ 
or 2,  or $\cos 2\theta_1=\cos 2\theta_2=0$. These cases correspond to 
$A(t_{res})=0$.}.  
Another possible realization of the parametric resonance condition is when 
neither of the terms on the right hand side of eq.~(\ref{res2}) vanishes
but they exactly cancel each other. 

\begin{samepage}
\begin{figure}
\setlength{\unitlength}{1cm}
\begin{center}
\vspace{-1.2cm}
\epsfig{file=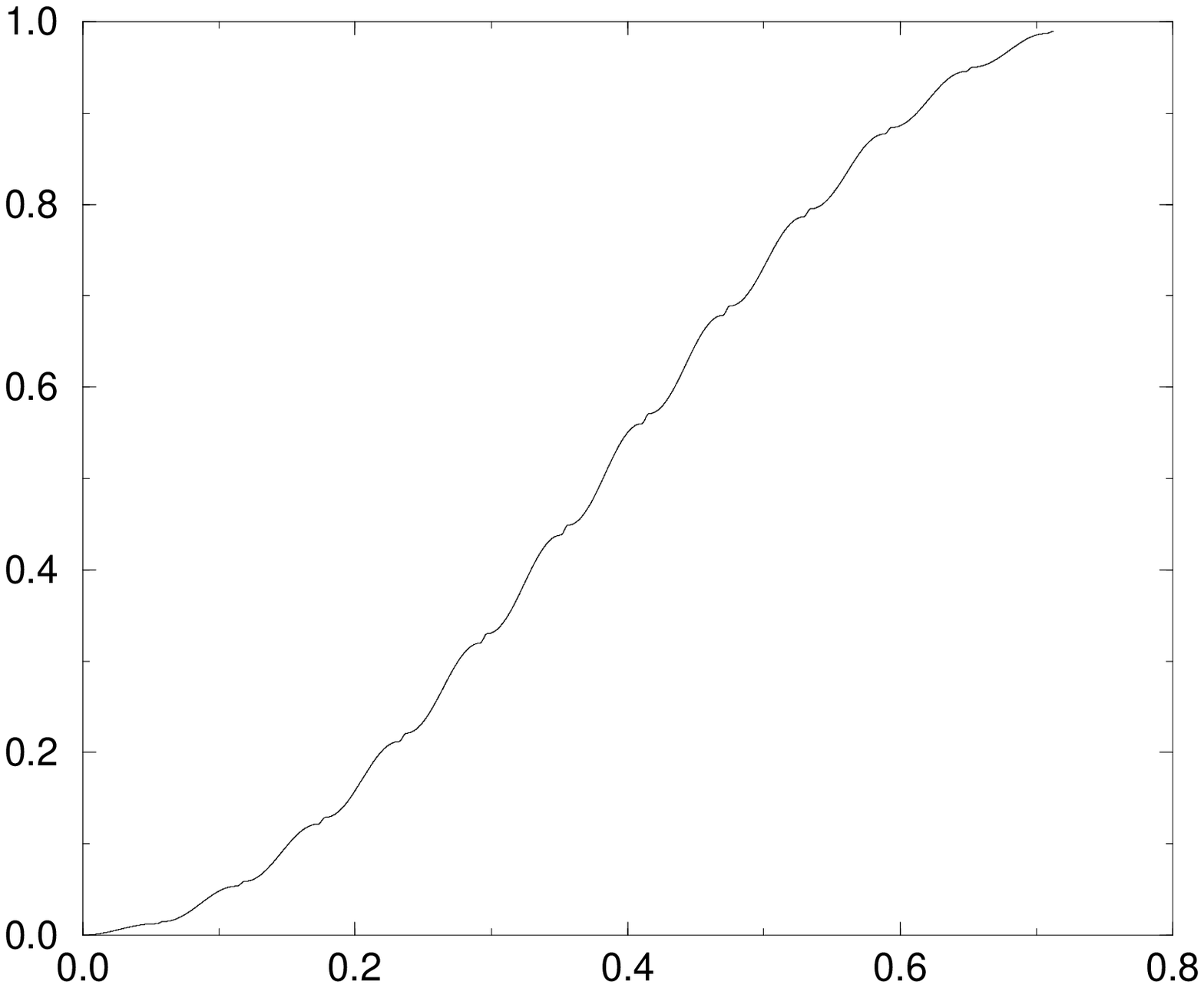,width=12cm}
\end{center}
\vspace{-1.4cm}   
\caption{\label{f3}
\small 
Same as in fig. 1 but for $\delta=10^{-12}$ eV, 
$V_2=10^{-11}$ eV, $T_1=5.4\times 10^{-2}$, $T_2=5.4\times 10^{-3}$.} 
\begin{center}
\epsfig{file=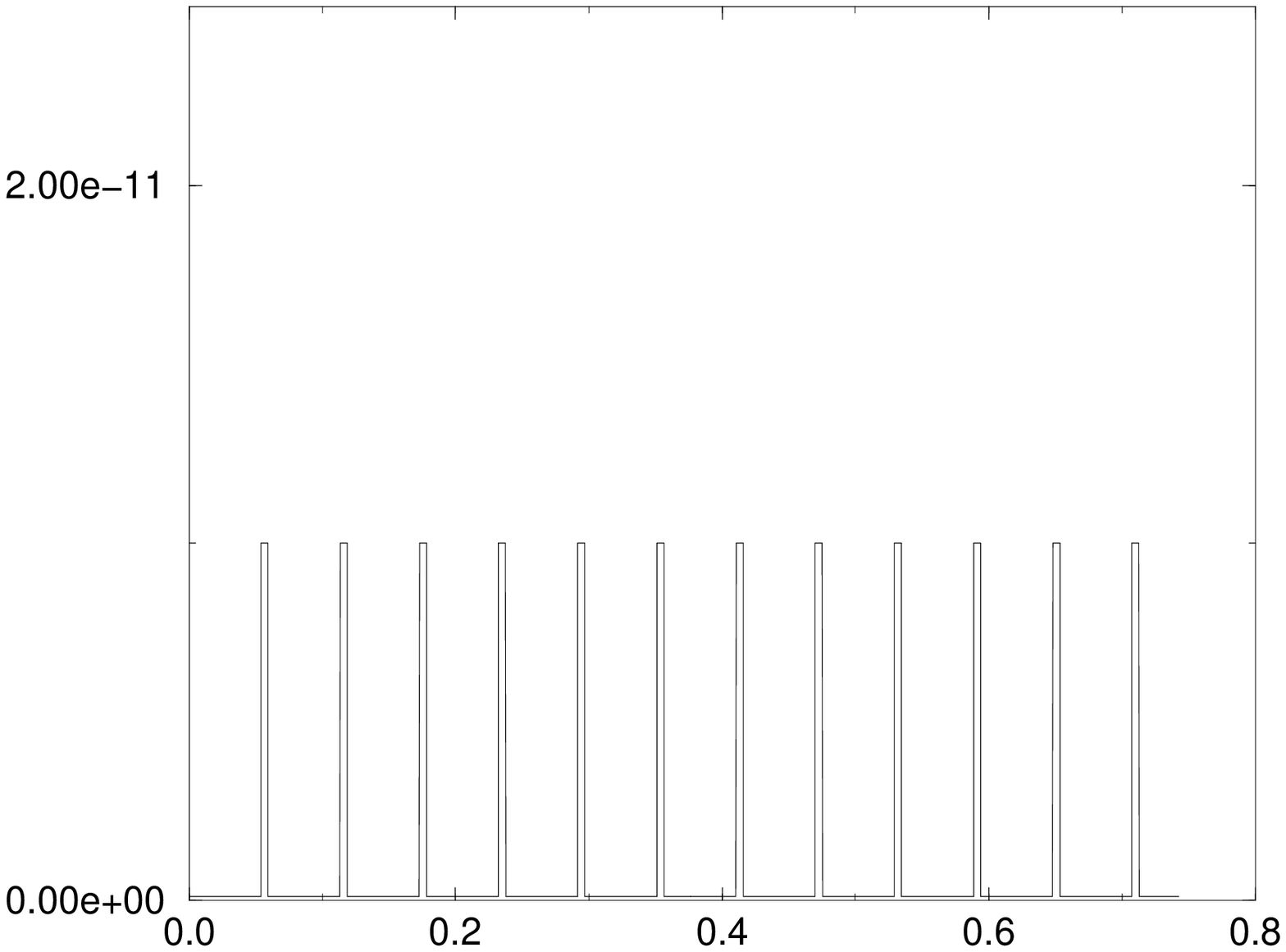,width=12cm}
\end{center}
\vspace{-1.4cm}   
\caption{\label{f4}
\small 
Coordinate dependence of the matter-induced neutrino potential 
for the case shown in fig. 3.}
\end{figure}
\end{samepage}  

We shall consider now the realization (\ref{rescond}) of the parametric 
resonance condition (\ref{res2}) (the second realization will be 
illustrated by a numerical example in fig. 7).  
At the resonance, the transition probability for the evolution over $k$
periods of density modulation takes a simple form 
\be
P(\nu_a\to \nu_b,~t=kT)=\sin^2 [k(2\theta_2-2\theta_1)] \,.
\label{probb1}
\ee
Let us first assume that the densities $N_1$, $N_2$ are either both below
the MSW resonance density $N_{MSW}$, which is determined from 
$G_F N_{MSW}/\sqrt{2}=\cos 2\theta_0 \,\delta$, or they are both above it. 
This means that the mixing angles in matter $\theta_{1,2}$ satisfy 
$\theta_{1,2}<\pi/4$ or $\theta_{1,2} > \pi/4$, respectively. It is easy
to see that in this case the difference $2\theta_2-2\theta_1$ is 
always farther away from $\pi/2$ than either $2\theta_1$ or
$2\theta_2$. Therefore in this case the transition probability
for evolution over one period cannot exceed the maximal transition 
probabilities in matter of constant density equal to either $N_1$ or
$N_2$, namely, $\sin^2 2\theta_1$ or $\sin^2 2\theta_2$. 
However, the parametric resonance does lead to an important gain.
In a medium of constant density $N_i$ the transition probability can never 
exceed $\sin^2 2\theta_i$, no matter how long the distance that neutrinos 
travel. On the contrary, in the matter with ``castle wall'' density
profile, if the parametric resonance conditions (\ref{rescond}) are
satisfied, the transition probability can become large provided neutrinos 
travel a large enough distance. 
It can be seen from (\ref{probb1}) that the transition probability can
become quite sizeable even for small $\sin^2 2\theta_1$ and 
$\sin^2 2\theta_2$ (i.e., in terms of the parameters of the
Hamiltonian (\ref{H1}), for small $max\{B^2/(A(t)^2+B^2)\}$).
This is illustrated in figs. 2 and 3 for the case $N_1,N_2<N_{MSW}$ (the
transition probability in the case $N_1,N_2>N_{MSW}$ has a similar behavior). 
The number of periods neutrinos have to pass in order to experience a complete 
(or almost complete) conversion is 
\be
k\simeq \frac{\pi}{4(\theta_1-\theta_2)}\,.
\label{number}
\ee

It is instructive to consider the limit of small density variations, 
$|N_1-N_2|\ll N_1$. In terms of the analogy with a pendulum with vertically 
oscillating point of support, it corresponds to the limit of the small 
amplitude of these vertical oscillations. In this limit $\theta_1\simeq
\theta_2$, and eq. (\ref{res2}) reduces to $\sin(\phi_1+\phi_2)=0$, or 
$\phi_1+\phi_2=k\pi$. Since $\phi_i=\omega_i T_i$, this condition can be
written as 
\be
\Omega\equiv \frac{2\pi}{T}=\frac{2\omega}{k}\,, \quad\quad
{\rm where}~~~~
\omega \equiv \omega_1 \frac{T_1}{T}+\omega_2 \frac{T_2}{T}\,.
\label{res3}
\ee
This coincides with the familiar parametric resonance condition in the
case of the small-amplitude variations of the parameter of the system, 
eq. (\ref{res1}). It is important to notice that the condition (\ref{res3}) 
does not depend on the amplitude of the density modulation, $N_1-N_2$. This
illustrates the point that we emphasized in sec. 3 -- the parametric
resonance can occur even for arbitrarily small amplitude of the variations 
of the parameters of the system. Of course, the smaller this amplitude, 
the longer the evolution time for the total conversion. 

Consider now the case $N_1<N_{MSW}<N_2$ ($\theta_1<\pi/4<\theta_2$). 
The transition probability over $n$ periods at the parametric resonance 
is again given by eq. (\ref{probb1}). However in this case, for $\theta_2 >
\pi/4 +\theta_1/2$ (which is always satisfied for small mixing in matter), 
one has $\sin^2 (2\theta_2-2\theta_1) > \sin^2 2\theta_1,\, \sin^2
2\theta_2$. This means that {\em even for the time interval equal to one 
period of matter density modulation the transition probability exceeds the 
maximal probabilities of oscillations in matter of constant densities
$N_1$ and $N_2$.} 
The case $N_1<N_{MSW}<N_2$ is illustrated in figs. 3, 6 and 7.  

Figs. 3 and 4 show the importance of the phase relationships in the case
of the parametric resonance.  
In these figures the coordinate dependence of the transition probability and 
matter density profile are shown for a specific case in which conditions 
(\ref{rescond}) are fulfilled. It can be seen from these figures that the
probability increase during the time intervals $T_2$, which correspond to the 
effective matter density $N_2$, is very small, and, in addition, in this case 
$T_2 \ll T_1$. One could therefore conclude that the evolution during these 
very narrow intervals is unimportant. However, this conclusion is wrong: if 
one removes the ``spikes'' in the matter density profile of fig. 4, i.e. 
replaces it by the profile $N(t)=N_1=const$, the resulting transition
probability will be very small at all times (fig. 5). 

\begin{figure}
\setlength{\unitlength}{1cm}
\begin{center}
\vspace{-1.2cm}
\epsfig{file=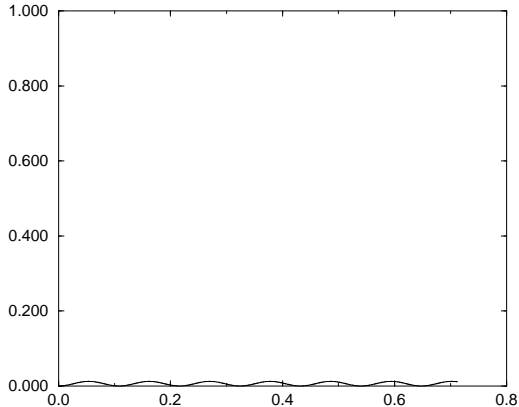,width=8cm}
\end{center}
\vspace{-1.4cm}   
\caption{\label{f5}
\small Same as in fig. 3 but for $V_2=V_1$ ($V(t)=V_1=const$).} 
\end{figure}

\begin{samepage}
\begin{figure}
\setlength{\unitlength}{1cm}
\begin{center}
\vspace{-1.2cm}
\epsfig{file=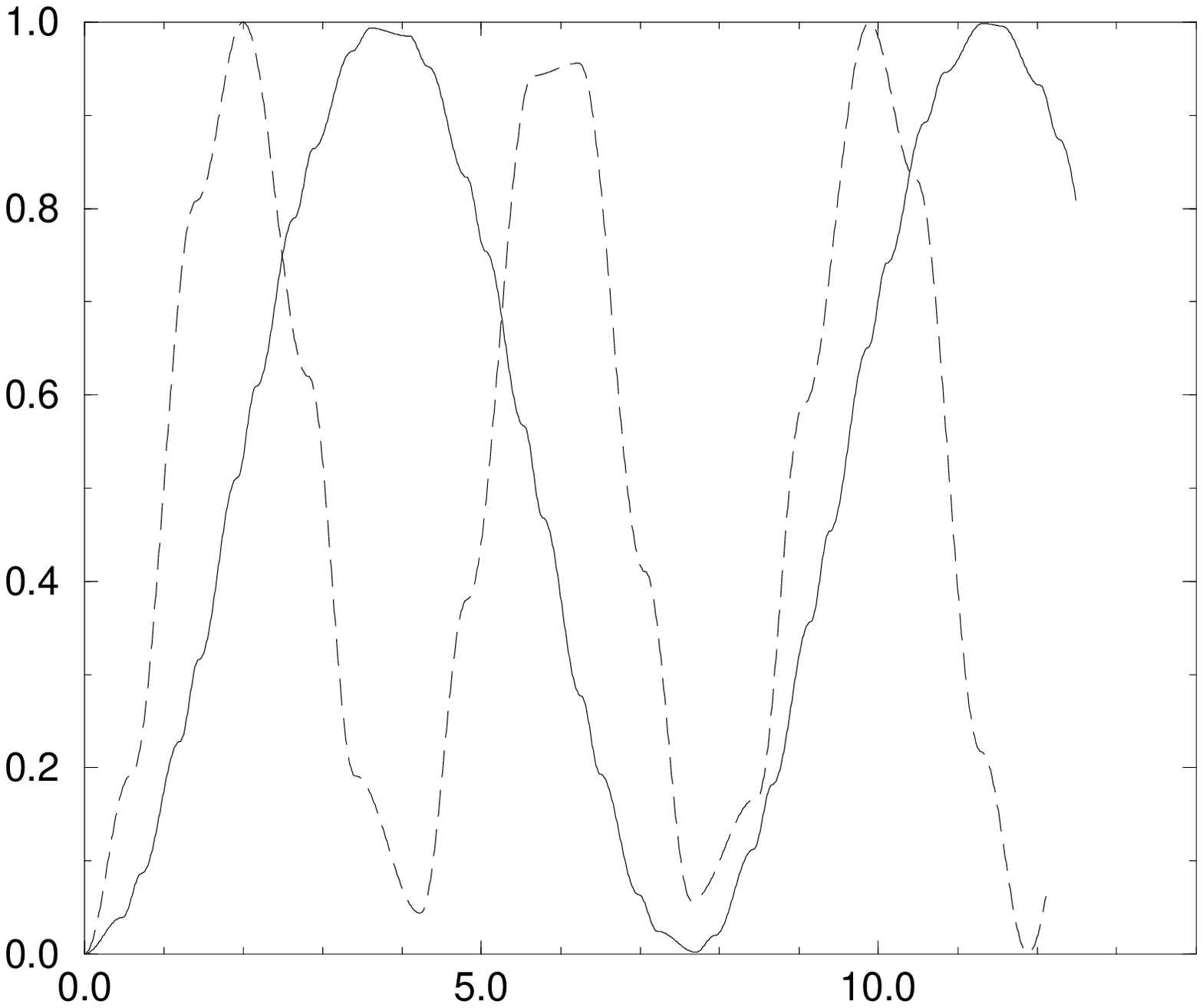,width=12cm}
\end{center}
\vspace{-1.4cm}   
\caption{\label{f6}
\small Solid curve: 
Coordinate dependence of the transition probability $P$ for the case of total 
conversion over 5 periods of density modulation (10 layers). Dashed curve: 
the same for the case of total conversion over 3 layers. The kinks correspond 
to the borders of the layers of different densities. The curves were
plotted for the realization (\ref{rescond}) ($c_1 = c_2 = 0$) of the
parametric resonance condition. 
}
\vspace{-0.4cm}   
\begin{center}
\epsfig{file=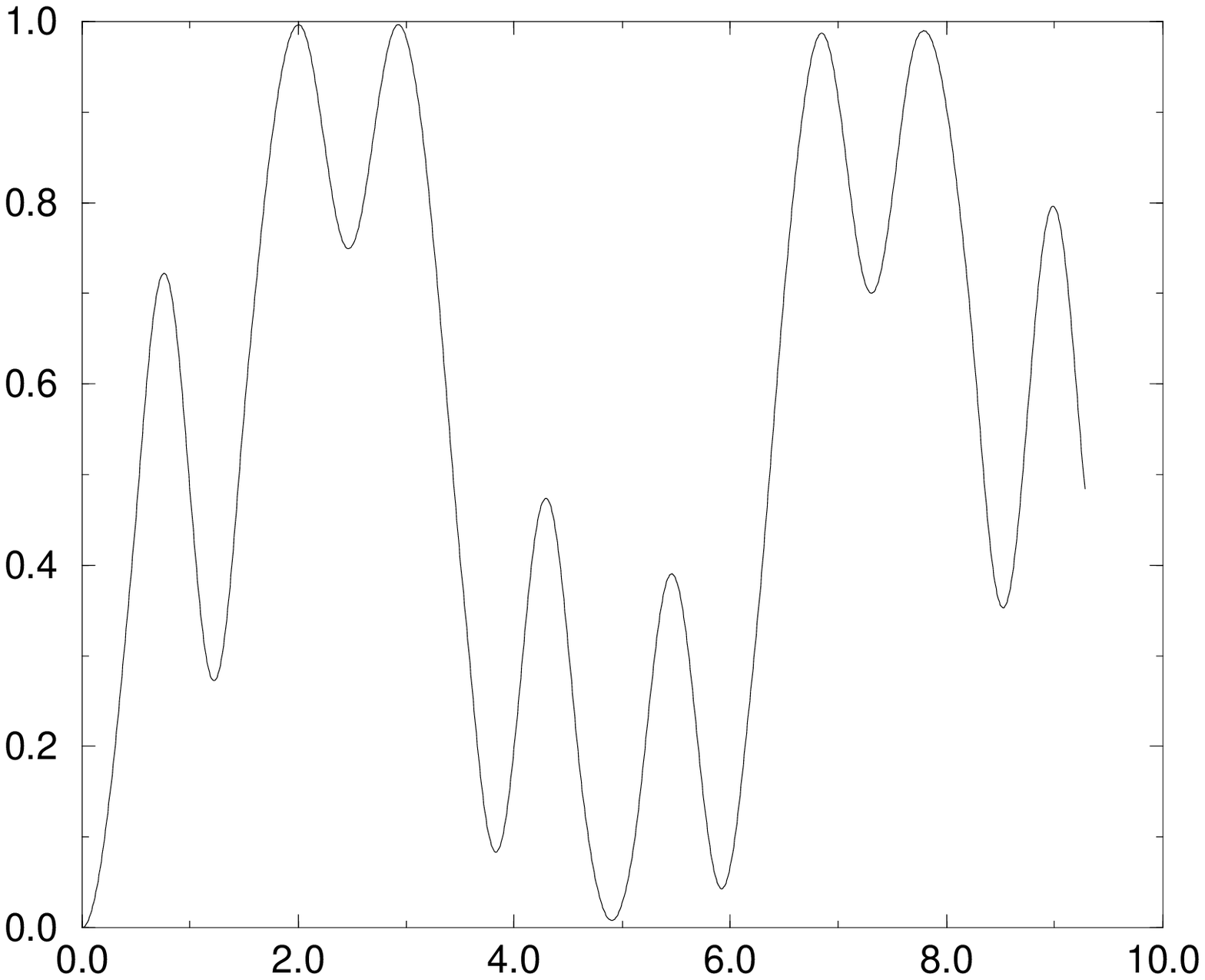,width=12cm}
\end{center}
\vspace{-1.4cm}   
\caption{\label{f7}
\small 
Same as in fig. 6 but for a case when the parametric resonance condition
is realized through the cancellation of the two terms in eq. (\ref{res1}); 
total conversion is achieved over 3 layers. 
}
\end{figure}
\end{samepage}  

Further examples of the parametrically enhanced neutrino oscillations can 
be found in figs. 6 and 7. Figures 1, 3 and 6 correspond to the
realization (\ref{rescond}) ($c_1=c_2=0$) of the parametric resonance
condition (\ref{res2}); fig. 7 illustrates the realization in which the two 
terms in $X_3$ cancel each other.

\section{Parametric resonance in neutrino oscillations in the earth}
\subsection{Evolution of oscillating neutrinos in the earth}

The earth consists of two main structures, the mantle and the core, which 
can to a very good approximation be considered as layers of
constant density. We shall consider neutrino oscillations in the earth in 
this two-layer approximation. Neutrinos coming to the detector from the 
lower hemisphere of the earth at zenith angles $\Theta$ in the range 
$\cos \Theta=(-1) \div (-0.837)$ (nadir angle $\Theta_n \equiv 180^\circ - 
\Theta \le 33.17^\circ$) traverse the earth's mantle, core and then again 
mantle, i.e. three layers of constant density with the third layer being 
identical to the first one. Therefore such neutrinos experience a periodic 
``castle wall'' potential, and their oscillations can be parametrically 
enhanced. Although the neutrinos propagate in this case only through three 
layers (``1.5 periods'' of density modulation), the parametric enhancement 
of the transition probability can be very strong. 

The evolution matrix in this case is $U=U_{1} U_{2} U_{1}$. It can be
parametrized in a form similar to that in eq. (\ref{UT2}): 
\be
U=Z-i\mbox{\boldmath $\sigma$}{\bf W}\,, \quad\quad Z^2+{\bf W}^2=1\,.
\label{U3}
\ee
The matrix $U$ describes the evolution of an arbitrary initial 
state and therefore contains all the information relevant for neutrino 
oscillations.  In particular, the probabilities of the neutrino flavor 
oscillations $P$ and of $\nu_2 \leftrightarrow \nu_{e}$ oscillations $P_{2e}$ 
(relevant for the oscillations of solar and supernova neutrinos inside
the earth) are given by \cite{Akh2} 
\footnote{Different but equivalent expressions can also be found in 
\cite{Min,Nic,P1}.} 
\be
P=W_1^2+W_2^2\,,\quad\quad 
P_{2e}=\sin^2 \theta_0+W_1(W_1 \cos 2\theta_0+W_3 \sin 2\theta_0)\,.
\label{prob}
\ee
Equivalently, the probability $P$ can be described by eqs. (\ref{prob1}) -- 
(\ref{varphi}) with $k=1$. 

We have now to identify the effective densities $N_1$ and $N_2$ with the
average matter densities $N_m$ and $N_c$ in the earth's mantle and core, 
respectively; similarly, we change the notation $V_{1,2}\to V_{m,c}$,
$\phi_{1,2}\to \phi_{m,c}$ and $\theta_{1,2}\to \theta_{m,c}$. 
 
In the two-layer approximation, the parameters $Z$, ${\bf W}$ have a very
simple form \cite{Akh2}: 
\be 
Z=2 \cos\phi_m \,Y-\cos\phi_c \,,
\label{Z}
\ee
\be
{\bf W}=\left(2\sin\phi_m \sin 2\theta_m\,Y+\sin\phi_c\sin 2\theta_c \,,
~0\,,~-\left(2 \sin\phi_m \cos 2\theta_m \,Y + \sin\phi_c \cos 2\theta_c
\right)\right) \,.
\label{W}
\ee
Here the vector ${\bf W}$ was written in components; the parameter 
$Y$ was defined in (\ref{Y}). 
If the parametric resonance condition (\ref{res2}) is satisfied through
the realization (\ref{rescond}), the neutrino flavor transition
probability takes the value \cite{LS,LMS}   
\be
P=\sin^2 (2\theta_c-4\theta_m) \,,
\label{prob2}
\ee
whereas the probability of the $\nu_2 \leftrightarrow \nu_{e}$ transitions is 
\cite{P1} 
\be
P_{2e} = \sin^2 (2\theta_c-4\theta_m+\theta_0) \,.
\label{prob3}
\ee
These probabilities can be close to unity (the arguments of the sines 
close to $\pi/2$) even if the amplitudes of neutrino oscillations in the 
mantle, $\sin^2 2\theta_m$, and in the core, $\sin^2 2\theta_c$, are rather 
small. This can happen if the neutrino energy lies in the range $E_c < E < 
E_m$, where $E_m$ and $E_c$ are the values of the energy that correspond to 
the MSW resonance in the mantle and in the core of the earth. 
This condition is equivalent to $N_m<N_{MSW}<N_c$.  
In the case of small mixing angle MSW solution of the solar neutrino
problem, $\sin^2 2\theta_0 < 10^{-2}$, and $P_{2e}$ practically coincides 
with $P$ unless both probabilities are very small.  

The trajectories of neutrinos traversing the earth are determined by their 
nadir angle $\Theta_n=180^\circ-\Theta$. The distances $R_m$ and $R_c$
that neutrinos travel in the mantle (each layer) and in the core are given by  
\be
R_m=R\left(\cos\Theta_n-\sqrt{r^2/R^2-\sin^2\Theta_n}\,\right)\,,
\quad\quad R_c=2R \,\sqrt{r^2/R^2-\sin^2\Theta_n}\,.
\label{RmRc}
\ee
Here $R=6371$ km is the earth's radius and $r=3486$ km is the radius of the 
core. The matter density in the mantle of the earth ranges from 2.7 $g/cm^3$ 
at the surface to 5.5 $g/cm^3$ at the bottom, and that in the core ranges
from 9.9 to 12.5 $g/cm^3$ (see, e.g., \cite{Stacey}). The electron number 
fraction $Y_e$ is close to 1/2 both in the mantle and in the core.
Taking the average matter densities in the mantle and core to be 4.5 and 11.5
$g/cm^2$ respectively, one finds for the $\nu_e\leftrightarrow \nu_{\mu,\tau}$
oscillations involving only active neutrinos the following values of $V_m$
and $V_c$: $V_m=8.58\times 10^{-14}$ eV, $V_c=2.19\times 10^{-13}$ eV.
For transitions involving sterile neutrinos $\nu_e\leftrightarrow
\nu_{s}$ and $\nu_{\mu,\tau}\leftrightarrow \nu_s$, these parameters
are a factor of two smaller.

\subsection{Parametric resonance conditions for neutrino oscillations in
the earth}

If the parametric resonance conditions (\ref{rescond}) are satisfied, strong 
parametric enhancement of the oscillations of core crossing neutrinos in the 
earth can occur \cite{LS,LMS,P1,Akh2,ADLS,CMP}, see fig. 8.   
In some cases, condition (\ref{phase}) can also be fulfilled, and the 
parametric resonance leads to a complete flavour conversion for neutrinos 
traversing the earth. 

We shall now discuss the resonance conditions (\ref{rescond}). The phases
$\phi_m$ and $\phi_c$ depend 
on the neutrino parameters $\Delta m^2$, $\theta_0$ and $E$ and also on the 
distances $R_m$ and $R_c$ that the neutrinos travel in the mantle and in the 
core. The path lengths $R_m$ and $R_c$ vary with the nadir angle; however,
as can be seen from (\ref{RmRc}), their changes are correlated and they cannot 
take arbitrary values. Therefore if for some values of the neutrino parameters 
a value of the nadir angle $\Theta_n$ exists for which, for example, the first
condition in eq. (\ref{rescond}) is satisfied, it is not obvious if at the same 
value of $\Theta_n$ the second condition will be satisfied as well. In other 
words, it is not clear if the realization (\ref{rescond}) of the parametric 
resonance condition (\ref{res2}) is possible for neutrino oscillations in the 
earth for at least one set of the neutrino parameters $\Delta m^2$, $\theta_0$ 
and $E$. However, as was shown in \cite{P1,Akh2}, not only the parametric 
resonance conditions (\ref{rescond}) are satisfied (or 
approximately satisfied) for a rather wide range of the nadir angles covering 
the earth's core, they are fulfilled for the ranges of neutrino parameters 
which are of interest for the neutrino oscillations solutions of the solar and
atmospheric neutrino problems. In particular, the conditions for the principal 
resonance ($k'=k''=0$) are satisfied to a good accuracy for $\sin^2 2\theta_0 
\aprle 0.1$, $\delta \simeq (1.1 \div 1.9)\times 10^{-13}$ eV$^2$, which 
includes the ranges relevant for the small mixing angle MSW solution of the 
solar neutrino problem and for the subdominant $\nu_\mu \leftrightarrow 
\nu_{e}$ and $\nu_e \leftrightarrow \nu_{\tau}$ oscillations of atmospheric 
neutrinos. 

\begin{figure}
\setlength{\unitlength}{1cm}
\begin{center}
\vspace{-1.2cm}
\epsfig{file=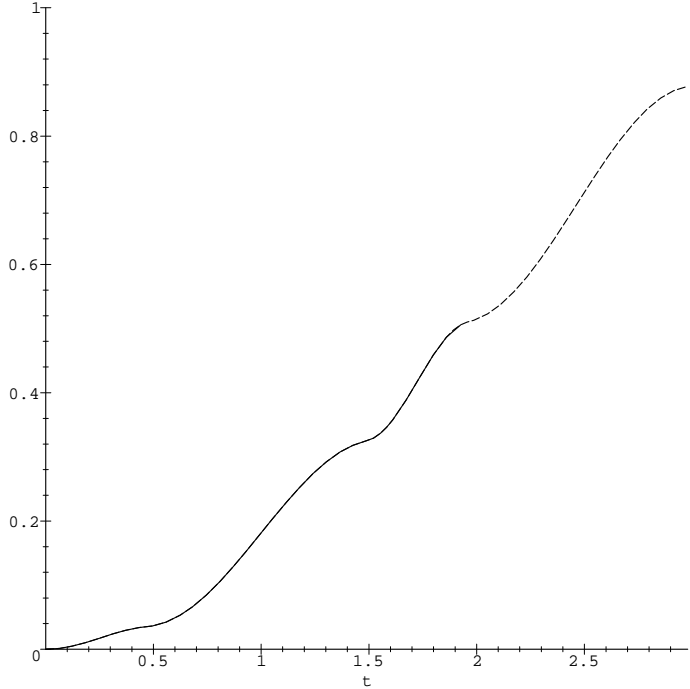,width=12cm}
\end{center}
\vspace{-1.4cm}   
\caption{\label{f8}
\small 
Solid curve: transition probability $P$ for 
$\nu_e\leftrightarrow \nu_{\mu,\tau}$ oscillations 
in the earth as a function of the distance $t$ (measured in units of
the earth's radius $R$) along the neutrino trajectory. 
$\delta\equiv \Delta m^2/4E =1.8\times 
10^{-13}$ eV, $\sin^2 2\theta_{0}=0.01$, $\Theta_n=11.5^\circ$. Dashed 
curve: the same for a hypothetical case of neutrino propagation over 
full two periods of density modulation ($t_{max}=2(R_m+R_c$)).
}
\end{figure}

The fact that the parametric resonance conditions (\ref{rescond}) can be
satisfied so well for neutrino oscillations in the earth is rather surprising. 
It is a consequence of a number of remarkable numerical coincidences. It has 
been known for some time that the potentials $V_m$ and $V_c$  corresponding to 
the matter densities in the mantle and core, the inverse radius of the 
earth $R^{-1}$, and typical values of $\delta\equiv \Delta m^2/4E$ of
interest for solar and atmospheric neutrinos, are all of the same order of
magnitude -- ($3\times 10^{-14}$ -- $3\times 10^{-13}$) eV (see, e.g., 
\cite{GKR,LS,LiLu}). It is this surprising coincidence that makes appreciable 
earth matter effects on the 
oscillations of solar and atmospheric neutrinos possible. However, for the
parametric resonance to take place, a coincidence by an order of magnitude is 
not sufficient: the conditions (\ref{rescond}) have to be satisfied at least 
within a 50\% accuracy \cite{Akh2}. This is exactly what takes place. In 
addition, in a wide range of the nadir angles $\Theta_n$, with changing 
$\Theta_n$ the value of the parameter $\delta$ at which the resonance 
conditions (\ref{rescond}) are satisfied slightly changes, but the fulfillment 
of these conditions is not destroyed. 

Even more surprising is the fact that the second realization of the parametric 
resonance condition (\ref{res2}), the one in which the two terms in $X_3$ 
cancel each other, is also possible for neutrino oscillations in the earth 
\cite{CP}. This requires a very subtle tuning between the values of
neutrino parameters $\theta_0$ and $\delta=\Delta m^2/4E$ on one hand and the 
effective matter densities $N_m$ and $N_c$ and neutrino pathlengths $R_m$ and 
$R_c$ in the mantle and in the core on the other. 
This looks very contrived, and yet turns out to be possible. Moreover, for a 
number of the values of the neutrino parameters, the condition (\ref{phase}) 
can also be fulfilled and a complete flavour conversion for neutrinos crossing 
the earth is is possible \cite{CP}.

\section{Discussion and conclusion}

We have reviewed the Floquet theory of linear differential equations 
with periodic coefficients and discussed its applications for neutrino 
oscillations in matter of periodically modulated density. In particular, 
in the case of two-flavour oscillations, we have 
shown that the evolution of the system takes a form of parametric 
oscillations -- modulated oscillations characterized by two periods, the 
period $T$ of density modulations and $\tau=(2\pi/\Phi)T$, where $\pm i\Phi$
are the characteristic exponents. We have also discussed the parametric 
resonance in neutrino oscillations and have shown that, irrespective of 
the shape of the periodic density modulations, the parametric resonance
condition is $X_3=0$, where $X_3$ is one of the parameters that determine
the monodromy matrix given in eq. (\ref{UT2}). We have also reviewed the 
exact solution in the case of 2-flavour neutrino oscillations in a matter
of periodic step-function density profile, obtained in \cite{Akh1, Akh2}, 
and discussed its relations with the Floquet theory. This solutions allows 
one to find explicit formulas for the parameters entering into the 
monodromy matrix, including the characteristic exponents and $X_3$. 
We discussed the implications of this exact solution for the oscillations
of neutrinos inside the earth, the density profile of which can to a very
good accuracy be approximated by a piece of the periodic step-function 
profile. We concentrated on possible parametric resonance effects in
neutrino oscillations in the earth. 

Besides being an interesting physical phenomenon, the parametric resonance in  
neutrino oscillations can provide us with an important additional information  
about neutrino properties. Therefore experimental observation of this effect 
would be of considerable interest. 
The prospects for experimental observation of the parametric resonance in 
oscillations of solar and atmospheric neutrinos traversing the earth were 
discussed in \cite{Akh3,Akh4}. To a large extent, these prospects depend 
on the values of some of neutrino parameters which are not well known yet.
The bottom line is that such an experimental observation is difficult but 
may be possible. 

Parametric enhancement may lead to noticeable effects in oscillations of
supernova neutrinos in the earth, resulting in characteristic distortions 
of the spectra of neutrinos crossing the earth's core \cite{DS}. A 
sufficiently accurate measurement of the supernova neutrino spectrum 
would, however, require a relatively close supernova ($L\le 10$ kpc).  

An interesting possibility to study the parametric effects in neutrino
oscillations would be very long baseline experiments with intense neutrino
beams produced at neutrino factories (for discussions of neutrino 
oscillations experiments at neutrino factories see, e.g., \cite{nufact}). 
For baselines larger than approximately 10700 km, neutrinos would cross the 
earth's core, and it would be possible to probe the parametric resonance 
effects in $\nu_e \leftrightarrow  \nu_{\mu(\tau)}$ oscillations. Notice,
however, that the present feasibility studies concentrate on relatively 
short baselines, a few thousand km \cite{nufact}. In addition, the expected 
average energies of neutrino beams ($E\ge 20$ GeV) in the currently discussed 
experiments are somewhat higher than what would be desirable in order to 
study the parametric effects. 

As we have seen, observing the parametric resonance in oscillations of solar,  
atmospheric or supernova neutrinos in the earth or in experiments with 
neutrino factories is not an easy task. Can one create the necessary matter 
density profile and observe the parametric resonance in neutrino oscillations 
in the laboratory (i.e. short-baseline) experiments? Unfortunately, the answer 
to this question seems to be negative: this would require either too long a 
baseline or neutrino propagation in a matter of too high a density (see 
\cite{Akh3,Akh4} for details). One can conclude that the sole presently known 
object where the parametric resonance in neutrino oscillations can take place 
is our planet, as was first pointed out in \cite{LS,LMS}. 

This work was supported by Funda\c{c}\~ao para a Ci\^encia e a Tecnologia 
through the grant PRAXIS XXI/BCC/16414/98.

\end{document}